\documentstyle[multicol,aps,epsf]{revtex}
\begin{document}
\draft
\title{Dynamics and stress in gravity driven granular flow}
\author{Colin Denniston$^{1}$ and Hao Li$^{2}$}
\address{$^{1}$ Dept. of Physics, Theoretical Physics,
University of Oxford, 1 Keble Road, Oxford OX1 3NP}
\address{$^{2}$ Center for Studies in Physics and Biology,
Rockefeller University, 1230 York Ave, New York, NY 10021}

\date{\today}
\maketitle

\begin{abstract}
 We study, using simulations, the steady-state flow of dry sand driven
 by gravity in two-dimensions.  An investigation of the microscopic
 grain dynamics reveals that grains remain separated but with a
 power-law distribution of distances and times between collisions.  
 While there are large random grain velocities, many of these
 fluctuations are correlated across the system and local 
 rearrangements are very slow. Stresses in the system are almost
 entirely transfered by collisions and the structure of the stress
 tensor comes almost entirely from a bias in the directions in which
 collisions occur.
\end{abstract}

\begin{multicols}{2}
Despite being the {\it raison d'\^{e}tre} for one of the oldest
scientific instruments, an understanding of the dynamics of sand
grains in the hourglass remains a mystery which still leaves much to
be explored.  Recent research into granular material has followed a
number of different tracks \cite{review}.  The initial state of a
diffuse granular gas with velocity fluctuations (about some mean flow
velocity $\langle{\bf v}\rangle$) drawn from a Maxwell-Boltzmann
distribution with characteristic ``granular'' temperature $T$, can be
described with fluid-mechanical equations using the ideas and
machinery of kinetic theory \cite{Haff,JenkinsSavage,JenkinsRichman}.
Yet, left to themselves, such systems are unstable to the formation of
high density clusters \cite{GoldhirschZanetti,McNamaraYoung}.  As the
system cools, due to the kinetic energy of the grains being dissipated
in the inelastic collisions, correlations in the density and spatial
velocity build up and hydrodynamic descriptions based on molecular
chaos (mean field assumption) break down \cite{NoijeErnstetal,Haoetal,Esipov}.
If inelastic collapse to a static state is to be avoided, the system
must be constantly supplied with energy.

Here, we consider inelastic hard spheres under the influence of
gravity in a cylindrical container (chute flow) with a finite
probability of reflection $p$, at the bottom (equivalent to a sieve in an
experiment).  The dynamics conserve momentum and the energy loss in a
collision is proportional to $1-\mu^2$ where $\mu$ is the coefficient
of normal restitution.  [The grain dynamics are the same as those used in
\cite{GoldhirschZanetti,McNamaraYoung,NoijeErnstetal,Grossman} and
others].  Due to the flow restriction imposed by the sieve at the
bottom, the density 
is very high (around $90\%$ of that of a closed packed configuration).  
The mean velocity profiles are very similar to those found in quasi-static
flows \cite{ShearZoneExp,PouliquenGutfraind}, a fairly flat region in
the middle with exponential boundary layers, or shear zones at the
edges.  However, as there is no friction in the system and the grains
remain separated, a different explanation is required here.  
We believe the ability of the system to withstand shear in this regime is
due to the very long time scales related to the slow glass-like
rearrangement of grains.  Diffusion is extremely limited, something
that is reflected in the finding that both the distribution of
distances and times between collisions is a power law.  
In addition, we find that the stress tensor is not simply related to
gradients in the density or velocity fields, as suggested by hydrodynamic-type
descriptions of more diffuse granular flows\cite{Haff,Savage}, but
obtains its structure primarily from an anisotropic distribution of
collision impact directions.  The structure of the stress tensor is
thus found to be similar to recently proposed models of stress
propagation in static sand in which the direction of the
characteristics for stress propagation is fixed \cite{Catesetal}. This
gives insight into how such models of the static case may be obtained
from the dynamics which were involved in the creation of static
configurations of grains.
Finally, we find large velocity fluctuations in the bulk regions which
are free from shear, in agreement with recent experiments
\cite{MenonDurian}.  This apparent contradiction with the more
accepted view that velocity fluctuations should be zero in the absence
of shear \cite{Haff,Savage} can be resolved when one takes into
account heat flow.  It is found that velocity fluctuations are
generated in the shear zones and are then conducted to the interior
region.  

\begin{figure}
\narrowtext
\centerline{\epsfxsize=3.1in
\epsffile{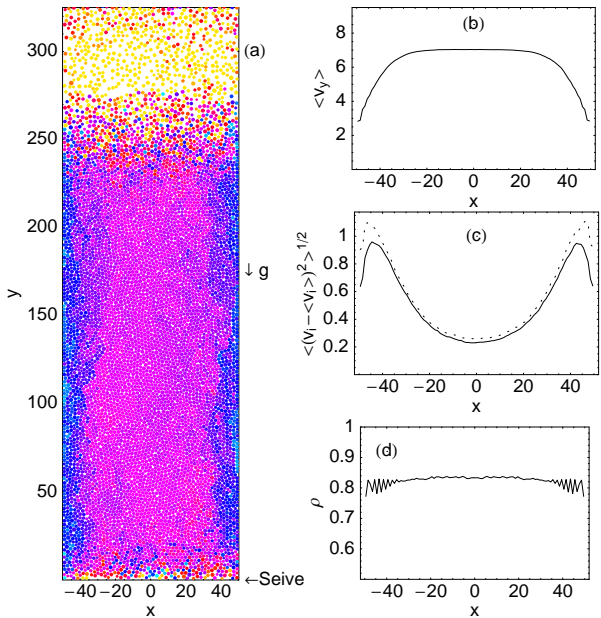}}
\vskip 0.1true cm
\caption{(a) Sand-flow in a 2-dimensional cylindrical geometry.  Gravity
  is in $y$-direction.  The color codes the $y$-velocity varying from
  yellow for fast to blue for slow.  Grains are circular with
  $10\%$ polydispersity, coefficient of restitution $\mu=0.95$, and
  the probability of reflection at the bottom is $p=0.5$.  Profiles
  across the chute at $y=100$ of: (b) Average
  vertical velocity $\langle v_y \rangle$;
  (c) Root mean square velocity fluctuations for the $x$ (solid) and
  $y$-components (dashed); (d) Average density $\rho$ as a function
  of $x$.}
\label{figure1}
\end{figure}

Figure \ref{figure1}(a) shows a snapshot of one of our 2-d simulations.
When a grain hits the bottom of the chute, it is reflected with
probability $p$ and escapes with probability $1-p$.  Whenever a grain
escapes from the bottom a new grain is placed at the top.  The grains
are polydisperse with the diameter of the $i$th grain being
$d_i=2+\epsilon_d \delta_i$ where the $\delta_i$ are independently
distributed Gaussian random variables with unit standard deviation,
and $\epsilon_d$ typically around $0.1-0.2$.  This polydispersity breaks
up the long range crystalline order which would otherwise be present
(crystalline order still persists over a distance of 3-5 grains).
When two grains collide, the velocities after
collision $\dot{\bf r}_1'$ and $\dot{\bf r}_2'$, expressed in terms of
the velocities before collision, $\dot{\bf r}_1$ and $\dot{\bf r}_2$, are
\begin{equation}
\left( \matrix{\dot{\bf r}_1' \cr \dot{\bf r}_2' \cr} \right)=
\left( \matrix{\dot{\bf r}_1 \cr \dot{\bf r}_2 \cr} \right)+
{1\over 2}(1+\mu)\left( \matrix{-1 & 1 \cr 1 & -1 \cr}\right)
\left( \matrix{\dot{\bf r}_1 \cdot {\bf q}  \cr \dot{\bf r}_2
\cdot {\bf q} \cr} \right){\bf q},
\label{collide}
\end{equation}
where ${\bf q}=({\bf r}_2-{\bf r}_1)/|{\bf r}_2-{\bf r}_1|$, and $\mu$
is the coefficient of restitution (typically $\mu=0.95$ or $\mu=0.9$
in our simulations).  While energy is lost to the internal energy of
individual grains, momentum is conserved. For $0<p<1$, a steady state
situation is achieved where a dense column of granular material flows
down the chute with a constant (in time) average velocity.  Within
this column, grains
\noindent {\it remain separated}, flying from collision
to collision with fairly large velocities.  The grains advance between
collisions following Newton's equations of motion.  Around $10^9$
collisions were tracked ($\sim 10^6$ per grain) with an additional
$10^8$ collisions discarded at the beginning to ensure we had reached
steady state conditions.  During this time the grains travel through
the system about 20 times. 
The width of the chute (50 or 100 radii) is comparable or larger than
the smallest dimension of situations studied experimentally in
3-dimensions in \cite{MenonDurian}.

If, as in the experiments, the pressure is to be independent of the
height of the column the side walls must in some way be able to
support the weight or absorb momentum in the vertical direction.  
In order to do this we make collisions with the side walls inelastic
in the tangential direction so that after colliding with a side wall,
the velocity $\dot{\bf r}'$ in terms of the velocity before the collision is
$\dot{\bf r}'=(-\dot{\bf r}_x,\mu \dot{\bf r}_y)$ (There is nothing
special about the choice of $\mu$ for the factor by which $\dot{\bf
r}_y$ is decreased).  One can think of this as a very simple model of
``soft'' walls (softer than the grains, e.g. Plexiglas walls and
brass balls) or an even simpler model of a rough surface.  As we
shall see, the boundary layers are exponentially small so that the
precise boundary conditions should not affect grain motion in the interior.

\begin{figure}
\narrowtext
\centerline{\epsfxsize=3.1in
\epsffile{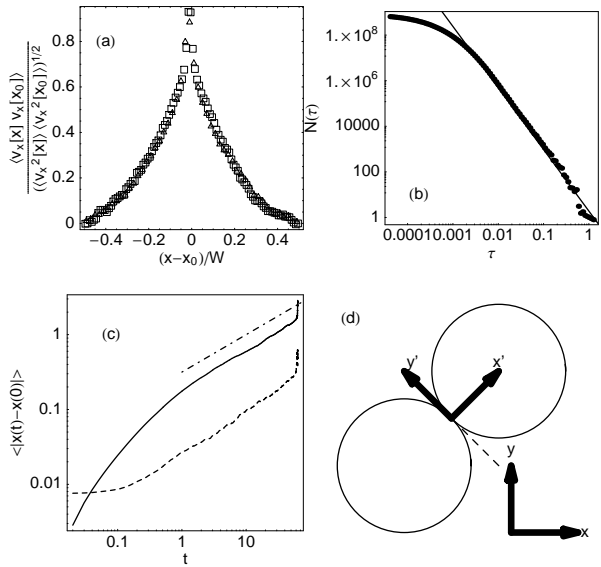}}
\vskip 0.1true cm
\caption{(a) Correlation function for the $x$ component of velocity shown as
  a function of the scaled variable $(x-x_0)/W$ where $W$ is
  the width of the system ($W=50$ for the triangles and $100$ for the
  squares) and measured from $(x_0,y_0)=(0,133)$, a point at
  the middle of the column.
  (b) Unnormalized probability $N(\tau)$ that a ball
  survives a time $\tau$ 
  without suffering a collision, as measured near
  the center of the column.  The straight line indicates the
  power law $\tau^{-2.75}$. (c) For a grain starting at
  $r(0)=(0,180)$, the average distance (solid line) a grain moves in
  the $x$ direction as a function of time; distance between two
  particles which started out as nearest neighbors (dashed) and; 
  power law $t^{1/2}$ expected for normal diffusion (dot-dashed).  
  In (b) and (c) data is from a system of width $50$, $p=0.5$, and $\mu=0.9$.
  (d) When two grains collide momentum is transferred in the $x'$
  direction across the intervening plane marked by the dashed
  line.}
\label{figure2}
\end{figure}

Figure~\ref{figure1}(b) plots the average vertical velocity in a cross-section
of the chute.  The velocity is flat in the central region,
implying an ability to withstand finite shear stresses, with
exponential boundary layers, or shear zones.  These profiles are very
similar to those seen in slow and quasi-static flows
\cite{Savage,ShearZoneExp,PouliquenGutfraind}.  However, the
explanation for quasi-static flows involves overcoming the yield
stress which arises from static friction \cite{PouliquenGutfraind}.
Here there is no friction between the grains and the grains remain
separated (and undergo fairly large velocity fluctuations,
Figure~\ref{figure1}(c)).  
The ability to withstand shear is less surprising
when you consider correlations in the velocity fluctuations.
Figure~\ref{figure2}(a) shows the velocity-velocity
correlation function for the $x$ component of velocity.  As can be
seen, the velocity correlation function scales with system size.  
As we shall see below, this correlated motion allows for the coherent
transfer of momentum across the system from collision to collision, thereby
resulting in some of the solid-like properties.

Before examining the process of stress transfer, it is worthwhile to
look at the underlying microscopic dynamics, and how these differ from
those in a normal fluid or solid material. In diffuse regions, the
mean free path can be a very useful quantity
to measure \cite{Grossman}.  Figure~\ref{figure2}(b) shows the
(unnormalized) probability $N(\tau)$ that a grain survives a time
$\tau$ without suffering a collision, measured near the center of the
dense column.  As can be readily seen, $N(\tau)$ has a power law
dependence on $\tau$: 
\begin{equation}
N(\tau) \sim \tau^{-\alpha}
\end{equation}
where $\alpha=2.75 \pm 0.05$.  There is a short-time cut-off 
where $N(\tau)\rightarrow constant$ as $\tau\rightarrow 0$.  In
addition, the distance traveled 
between collisions $l$ is also a power law with the same exponent.
This is not that surprising since the moments of
the velocity distributions appear to be convergent and thus $\tau$ and
$l$ are related by some typical velocity. The power law leads to a
typical mean free path comparable to the short distance cutoff given
by the inter-particle spacing.  This suggests that particles are
tightly confined by their neighbors. 

To examine this, it is instructive to look at the observed diffusion
of a grain as it travels through the system.
Figure~\ref{figure2}(c) shows the average distance traveled in the
$x$-direction by a single particle (solid line) in a time $t$
and the distance between two particles (dashed line)
which started out as nearest neighbors near the top of the column.
One sees that diffusion is very limited and that the two particles
typically remain
nearest neighbors throughout their lifetime in the column.  Even the
single particle motion is fairly limited moving only of order 1 radius
during it's lifetime in the column.  The single particle motion is not
diffusive (diffusive motion would have $\langle |x(t)-x(0)| \rangle \sim
t^{1/2}$) but appears to be approaching this in the long time limit.

Momentum conservation requires that
\begin{equation}
\partial_t(\rho v_i)+\partial_k(\rho v_i v_k)=\partial_k\sigma_{ik}+\rho g_i
\label{momenta}
\end{equation}
where $\sigma_{ik}$ is the stress tensor, and all $v_k$ refer to first
moments of the velocity distribution.  In our steady state situation
the left hand side is zero.  
There are two types
\noindent of contributions to the stress tensor.
The first contribution comes from
the diffusion of particles (By this we do not mean macroscopic
convection, the $\partial_k(\rho v_i v_k)$ term, but the microscopic
particle diffusion, present even in the case of no macroscopic flow).
This contribution turns out to be negligible as particles do not
diffuse around very much (Fig.~\ref{figure2}(c)).  Unlike the often
studied granular gas case, in the situation we study here the stress
is dominated by momentum transfer during collisions.

\begin{figure}
\narrowtext
\centerline{\epsfxsize=3.1in
\epsffile{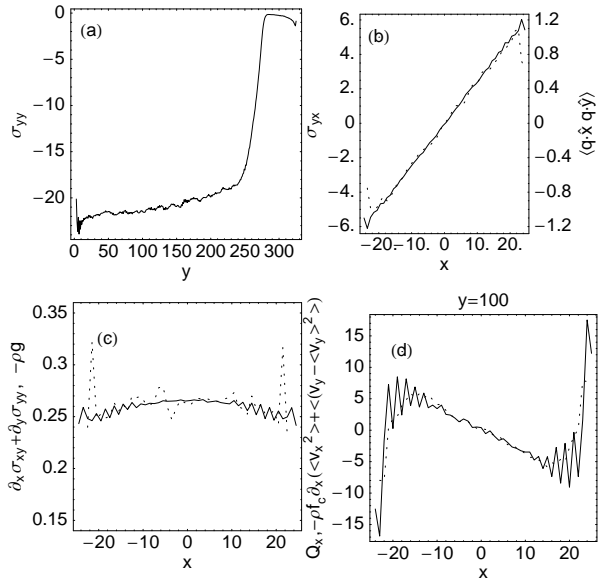}}
\vskip 0.1true cm
\caption{(a) Diagonal components of the stress tensor $\sigma_{yy}$
  computed from momentum transfer during collisions.  (b) Off-Diagonal
  component of the stress tensor, $\sigma_{yx}$ (solid line and left
  axes) and correlation between $x$ and $y$ components of a vector
  ${\bf q}$ drawn at every collision from the center of one grain to
  the center of the other grain (dashed line and right axes).  (c)
  Balancing of body force $\rho g$ (solid line) and 
  derivatives of stress tensor $(\partial_x \sigma_{xy}+\partial_y
  \sigma_{yy})$ (dotted line).  The larger fluctuations in $\rho g$
  near the edge are due to columnar ordering induced by the straight
  side walls.  Statistical error is comparable to the noise level in
  $(\partial_x \sigma_{xy}+\partial_y \sigma_{yy})$. Data is from the
  system of width $50$, $p=0.5$, and $\mu=0.9$. (d) Heat flow in
  $x$-direction $Q_x$ (solid) and $-\rho f_c {1\over 2}\partial_x\langle v^2
  \rangle$ (dotted).}
\label{figure3}
\end{figure}

Figure \ref{figure2}(d) shows two grains at the moment of collision.
The collision results in momentum being transferred in the $x'$
direction across the intervening plane marked by the dashed line in
the Figure. This gives a contribution to the stress tensor, measured
in the $(x',y')$ coordinate system of
\begin{equation}
d\sigma_{x'y'}=\left(
\matrix{{\bf p_c}\cdot \hat{\bf x}' & {\bf p_c}\cdot \hat{\bf y}' \cr
  0 & 0\cr} \right)
\end{equation}
where $p_c$ is the momentum transfered from grain $1$ to grain $2$.
Converting to $xy$-coordinates this is 
\begin{equation}
d\sigma_{xy}=\left(
\matrix{({\bf p_c}\cdot \hat{\bf x})(\hat{\bf x}\cdot \hat{\bf x}') & 
  ({\bf p_c}\cdot \hat{\bf y})(\hat{\bf x}\cdot \hat{\bf x}') \cr
  ({\bf p_c}\cdot \hat{\bf x})(\hat{\bf y}\cdot \hat{\bf x}') & 
  ({\bf p_c}\cdot \hat{\bf y})(\hat{\bf y}\cdot \hat{\bf x}')\cr}
\right).
\label{stresseq}
\end{equation}
The net contribution to the stress tensor is found by averaging over
time and dividing by the area over which the average is performed.
Figure~\ref{figure3}(a) and (b) shows $\sigma_{yy}$ and $\sigma_{xy}$ for one
particular case. 
Figure~\ref{figure3}(c) shows how,
within statistical errors, $\partial_k\sigma_{ik}+\rho g_i=0$ thus
satisfying Eq.(\ref{momenta}) (this relationship should be interpreted
in the probabilistic sense: It is true ``on average''). Interestingly,
most of the weight is supported by a shear stress, via $\partial_x \sigma_{xy}$
rather than the pressure gradient $\partial_y \sigma_{yy}\approx $
$ 0.014$.  Effectively, the column
behaves like a solid block sliding down a tube with friction at the
walls supporting the force due to gravity.  However, the system has
solid properties only in the sense of a glass, as rearrangements do
occur, albeit on longer time scales. 

Given that things are not isotropic, it is important to see
where the structure of the stress tensor comes from.  It can arise
from both gradients in velocities or from an anisotropic distribution
of collision directions. If
at every collision one draws a vector from the center of one grain to
the center of the other grain and then averages, one finds that there
is a definite bias in the collision directions
(Figure~\ref{figure3}(b),right axes).  This is consistent with the collision
``chains'' observed in the simulations and previously by other workers
\cite{McNamaraYoung,Esipov}. This 
bias in the collision directions implies a preferred direction to the
collision chains.  To clarify, we put the momentum transfer from
Eq.(\ref{collide}) into the stress tensor from Eq.(\ref{stresseq}) to get 
\begin{eqnarray}
\sigma_{ij}&=& {1 \over t}\sum_{collisions} -{1\over 2} (1+\mu)({\bf \dot r_1}-{\bf
  \dot r_2})\cdot {\bf \hat q} \times \nonumber\\
& & \qquad
\left(
\matrix{({\bf \hat q}\cdot {\bf \hat x})^2 & 
  ({\bf \hat q}\cdot {\bf \hat x})({\bf \hat q}\cdot {\bf \hat y}) \cr
  ({\bf \hat q}\cdot {\bf \hat y})({\bf \hat q}\cdot {\bf \hat x}) & 
  ({\bf \hat q}\cdot {\bf \hat y})^2\cr}
\right),\nonumber\\
& \approx & -{1 \over 2}(1+\mu)f_c \langle ({\bf \dot r_1}-{\bf
  \dot r_2})\cdot {\bf \hat q} \rangle \times \nonumber\\
& & \qquad
\left(
\matrix{\langle({\bf \hat q}\cdot {\bf \hat x})^2\rangle & 
  \langle({\bf \hat q}\cdot {\bf \hat x})({\bf \hat q}\cdot {\bf \hat y})\rangle \cr
  \langle({\bf \hat q}\cdot {\bf \hat y})({\bf \hat q}\cdot {\bf \hat x})\rangle & 
  \langle({\bf \hat q}\cdot {\bf \hat y})^2\rangle\cr}
\right),
\end{eqnarray}
where the sum is over collisions in a (long) time interval $t$ and $f_c$
is the collision frequency.  In writing this we have made an
assumption, supported by the data, that the factor $({\bf \dot r_1}-{\bf
  \dot r_2})\cdot {\bf \hat q}$ can be separated from the factors in
the matrix when computing averages.  In fact, the factors in front of
the matrix appear to be nearly a constant, hence
the structure of the stress tensor comes almost entirely from the bias
in the directions in which the collisions occur (cf. Fig.~\ref{figure3}(b)).
Note that there is a striking similarity between the structure of the
stress tensor in our model and that of a static model recently
proposed in Ref.~\cite{Catesetal}.  In our case, the dynamics of flow
automatically drive the system to a ``jamming'' configuration where
force chains (as given by the preferred directions of collision) form
and the system acquires an ability to withstand shear stress.

Finally, we examine the energy balance.  As we saw in
Figure~\ref{figure1}(c), there are fairly large velocity fluctuations
throughout the flow.  This is in agreement with the experiments of
Menon and Durian \cite{MenonDurian}, but in apparent contrast to
hydrodynamic descriptions of granular flows  
\noindent \cite{Haff,Savage} which suggest that velocity fluctuations about the mean flow  velocity should 
be zero in the absence of shear.  A resolution to this apparent
conflict may be found by examining the energy flow.  For our steady
state situations, energy conservation requires
\begin{equation}
0 = -\nabla \cdot {\bf Q}+\rho {\bf g}\cdot {\bf v}+\nabla \cdot ({\bf
  \sigma}\cdot {\bf v})+I.
\label{energy}
\end{equation}
 From Eq. (\ref{collide}), the (kinetic) energy lost in a
single collision is  
$\delta E=-{1 \over 4}(1-\mu^2)(\dot{\bf r}_1 \cdot {\bf q}-\dot{\bf
  r}_2 \cdot {\bf q})^2.$
Averaging over the collisions that occur in a unit area per unit time gives the
dissipation $I$.  As the stress balances gravity, the work done
by the second and third terms cancel, except for a contribution from
$\sigma_{xy} \partial_x \langle v_y \rangle$ in the shear zones at the
boundaries.  This leaves the ``heat'' flow term 
$(\nabla \cdot {\bf  Q})$ to cancel the dissipation in the bulk.  One
must make a clear distinction here between ``energy flux'' and ``heat
flux'', something that has not been necessary in the diffuse regime
\cite{Grossman}.  From Eq. (\ref{collide}), the energy flux from one
grain to another in a collision is $\delta {\bf F}={1 \over 2}(1+\mu)[(\dot{\bf r}_1\cdot {\bf q})^2-(\dot{\bf r}_2 \cdot {\bf q})^2]{\bf q}$.  However,
part of this energy flux is related to the coherent transfer of
momentum (i.e. it is the work done by the stress tensor).  As a
result, the heat flux is the uncorrelated part of the energy flux, 
${\bf Q}={\bf F}+({\bf \sigma}\cdot {\bf v})$.
With this done, one can relate the heat flux ${\bf Q}$ to the gradient
of the granular temperature $T={1 \over 2}(\langle v_x^2
\rangle+\langle (v_y-\langle v_y\rangle)^2 \rangle)$ by 
${\bf Q}=-\kappa \nabla T$.  As demonstrated in Figure~\ref{figure3}(d), the
conductivity $\kappa=\rho f_c$, not a terribly surprising result
although different from what one finds in the diffuse regime.


In conclusion, we find that the dynamics of gravity driven flow within
the regime studied do not fit into previous descriptions
of granular flow in the diffuse regime.  While the model of
instantaneous collisions with only normal forces is not necessarily
general to all granular materials, much of the behavior, such as the
presence of velocity fluctuations in the absence of shear does agree
with experiments \cite{MenonDurian} on hard grains.  While a complete
theory requires more work, our results put fairly tight constraints on
any such theory as well as providing very helpful information as to what are
reasonable (and unreasonable) assumptions in its formulation. 

We thank C. Tang, P. Chaikin, N. Menon, D. Durian and R. Stinchcombe 
for discussions. This work was funded in part by EPSRC (U.K.) GR/K97783.

\end{multicols}
\end{document}